\documentclass{cimento}

\usepackage{graphicx}  % got figures? uncomment this
\usepackage{epstopdf}
\usepackage{cite}
\usepackage{amssymb}

\title{Discovery of ${\nu_{\tau}}$ Appearance and Recent Results from OPERA}
\author{T.~Fukuda
\from{toho}
\thanks{Now at F-lab, Nagoya University - Nagoya, Japan.  \\
\hspace{12pt} E-mail: tfukuda@flab.phys.nagoya-u.ac.jp}, 
{\rm{on behalf of the}} OPERA Collaboration
}
\instlist{\inst{toho} Department of Physics, Toho University - Funabashi, Japan
}
\PACSes{\PACSit{14.60.Pq}{Neutrino mass and mixing}
}

\begin{document}

\maketitle

\begin{abstract}
The OPERA experiment was designed to observe ${\nu_{\mu}}$ ${\rightarrow}$ ${\nu_{\tau}}$ oscillations through ${\nu_{\tau}}$ appearance at a baseline of 730 km in the CNGS beam. Newly developed emulsion analysis technology allows to measure ${\nu_e}$, ${\nu_{\mu}}$ and ${\nu_{\tau}}$ interactions with the largest emulsion detector ever made. OPERA has detected five ${\nu_{\tau}}$ candidates, allowing to reject the background-only hypothesis at 5.1 ${\sigma}$. The parameters for standard and non-standard oscillations are measured and constrained.

\end{abstract}

\section{Introduction}

In 1962, Maki, Nakagawa and Sakata proposed that oscillation may exist between massive neutrinos of different flavors \cite{bib1}. The Super-Kamiokande experiment measured a deficit in atmospheric ${\nu_{\mu}}$ due to their disappearance through the oscillation mechanism \cite{bib2}. OPERA \cite{bib3} is an experiment that aims at detecting the appearance of ${\nu_{\tau}}$ in an almost pure ${\nu_{\mu}}$ beam to confirm the neutrino flavor transition expected from ${\nu_{\mu}}$ disappearance results of Super-Kamiokande, MACRO \cite{bib4}, K2K \cite{bib5}, MINOS \cite{bib6}. In 2010, a first ${\nu_{\tau}}$ candidate event was observed in OPERA \cite{bib7}. Then Super-Kamiokande reported ${\nu_{\tau}}$ appearance in atmospheric neutrino data with statistical analysis in a background-dominated sample \cite{bib8} and T2K also reported their ${\nu_e}$ appearance result in a ${\nu_{\mu}}$ beam \cite{bib9}. OPERA verifies the ${\nu_{\mu}}$ ${\rightarrow}$ ${\nu_{\tau}}$ oscillation at atmospheric scale in ${\nu_{\tau}}$ appearance mode on an event-by-event basis and with an extremely high signal-to-noise ratio (S/N$\sim$10). 

Tau neutrinos were observed for the first time by detecting the ${\tau}$ leptons produced in ${\nu_{\tau}}$CC interactions with Emulsion Cloud Chambers (ECC) in the DONUT experiment \cite{bib10}. The path length of the ${\tau}$ leptons is very short (c${\tau}$=87 ${\mu}$m) thus requiring very high spatial resolution. Nuclear Emulsion is a special type of photographic film made of AgBr microcrystals interspersed in a gel matrix. A charged particle passing through such medium ionizes the crystals along its path and produces a latent image. Upon a chemical process, the development, a particle trajectory is materialized by a line of grains of metallic Ag (0.5-1 ${\mu}$m diameter). An industrial mass production of nuclear emulsion was done for the first time for the OPERA experiment \cite{bib11}.

\section{Detector and beamline} 

\begin{figure}[ht]
\begin{center}
\includegraphics[clip, width=13.0cm]{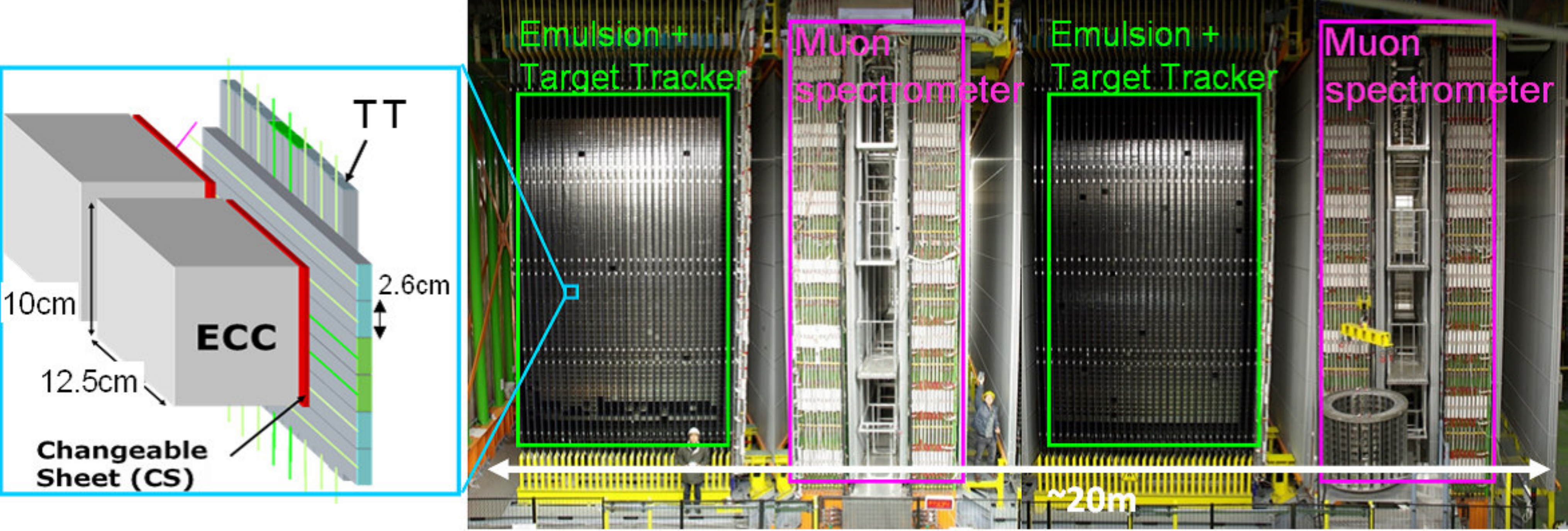}
\caption{The OPERA detector and its components.}
\label{fig-1}
\end{center}
\end{figure}

The OPERA detector \cite{bib12} is located at the Gran Sasso underground laboratory (LNGS) at a distance of 730 km from the neutrino source. The CERN CNGS ${\nu_{\mu}}$ beam \cite{bib13} had an average energy of 17 GeV, well above the ${\tau}$ lepton production threshold in ${\nu_{\tau}}$CC interactions. The rate of prompt ${\nu_{\tau}}$ is negligible. The oscillation probability is about 1.7${\%}$ at the most probable measured values of the oscillation parameters, ${\Delta}$\it{m}${^2 = 2.4}$ ${\times}$ $10^{-3}$ \rm{eV}${^2}$, sin${^22}$ ${\theta}$ = 1.0. From 2008 to 2012 the total amount of protons on target (pot) was 1.8 ${\times}$ $10^{20}$ pot, corresponding to 80${\%}$ of the design value. The OPERA detector is a hybrid detector composed of two identical Super Modules (SM) as shown in fig.\ref{fig-1}. The total target mass is 1.25 kton. In each SM, 75,000 ECC bricks were assembled into walls interleaved by two orthogonal planes of scintillator strips, the Target Tracker planes (TT), used to identify the bricks in which the interactions occurred. Each target is complemented by a spectrometer that identifies muons and measures their charge and momentum. The size of ECC brick is 12.8 cm ${\times}$ 10.2 cm ${\times}$ 7.9 cm. It is composed of 57 emulsion films (0.3 mm-thick) interleaved with 56 lead plates (1 mm-thick). A film has a 44${\mu}$m emulsion layer deposited on each side of a 205 ${\mu}$m plastic base. A separate box containing a pair of films, hereafter called Changeable Sheets or CS, is glued on the downstream face of each ECC brick in front of the next TT plane. The CS serve as interface between the brick and the TT, bringing the centimeter spatial resolution of the TT down to the ${\mu}$m level. 
%CS is key technology which is actualized the analysis of such a largest emulsion detector.
% Experimental requirements for OPERA are fulfilled with the CNGS beamline and the OPERA detector. The CNGS beamline supply the long baseline and high energy neutrino beam, then the OPERA detector supply the large target mass and high spatial resolution. 

\section{Neutrino event analysis}

The challenge in OPERA is the analysis of a large amount of emulsion films. The signal recorded by the electronic detectors, TT and muon spectrometers, is used to identify the bricks, most often one and up to 3, in which a neutrino interaction is likely to have occurred. Those bricks are extracted from the target by an automaton. CS are removed from their box, developed and analyzed, starting with the most probable brick. If tracks found on the CS are compatible with the electronic detector track, the brick is disassembled and its films developed. The tracks found in the CS are extrapolated to the most downstream emulsion film in ECC where they are searched for. They are then followed back from film to film up to the stopping point, where the tracks are not found in three consecutive films. A large volume around this point is scanned and analyzed, and the neutrino interaction vertex is reconstructed. Currently, out of the 16,879 triggered events, 7,041 neutrino vertices are localized in ECC, 6,682 events are fully analyzed.

The analysis procedure in the ECC had been well established in past experiments, CHORUS \cite{bib14} and DONUT \cite{bib15}. Given the large amount of emulsion scanning and analysis necessary for the OPERA experiment, newly developed emulsion analysis techniques played an essential role. The details will be described below.

\subsection{High speed scanning} \mbox{} S-UTS, ESS\\
In modern analysis of nuclear emulsions, the analysis speed has been dramatically improved due to the development of devices to automatically scan nuclear emulsions. The scanning speed of previous systems applied to DONUT was 1 cm$^2$/hour. Four S-UTS (Super Ultra Track Selector) \cite{bib16} with a scanning speed of 75 cm$^2$/hour and one of 20 cm$^2$/hour have been operated in Japan. The scanning speed of the ESS (European Scanning System) \cite{bib17} is 20 cm$^2$/hour and a total of 33 such systems have been run in Europe. Such significant improvement of scanning speed allow a large amount of emulsion data taking.

\subsection{Precision alignment} \mbox{} Compton Alignment\\
CS is constituted of two OPERA films and has four emulsion layers. A few tracks emerging from the ECC are expected to be recorded by the CS for each neutrino interaction. However $\sim$10$^8$ tracks on each layer are read out with scanning systems. The majority of them are not "real" or high momentum tracks from neutrino interactions, but so-called "fake" tracks which are due to low momentum electrons ($\sim$MeV) from radioactive source. By the CS analysis, it is possible to extract a few real tracks among such a huge number of fake tracks. The first step of fake track rejection is to require coincidence over 4 layers using 3 dimensional emulsion track data. By this requirement, background is limited to $\sim$10$^4$ tracks for CS.

New precise alignment method \cite{bib18} between CS films uses fake tracks (low momentum electrons) which cross the contacting surface between emulsion layers. Background electrons, called Compton electrons, are used for alignment instead of X-ray marks. The alignment accuracy of this new and conventional method is about 2 $\mu$m and 6-8 $\mu$m, respectively with an improvement by a factor of 3-4. The background level is reduced by a factor of 10 without any efficiency degradation.

\subsection{High S/N discrimination} \mbox{} Track Ranking\\
An ultimated S/N discrimination method of emulsion analysis is direct human eye check with a microscope. By human eye it is possible to recognize the scattered, not straight, fake tracks of low energy electron and distinguish real tracks from fake tracks. However human eye check takes about 1 min./track. In OPERA, it is feasible to apply such an eye check on 10-20 tracks on each CS at a rate of about 20 CSs per day. 

New S/N discrimination method \cite{bib19} is based on a likelihood analysis of the emulsion tracks. The evaluation parameter is focused on "the blackness" and "the linearity" of 4 layer tracks on CS. The blackness and the linearity of real tracks is higher than for fake tracks. 
%The blackness of fake tracks is thin since small number of grains constitute of fake tracks recognized as a line tracks by scanning systems because of scattering. Then the linearity of fake track for 4 emulsion layers is bad because they have no correlation on each layer and are a chance coincidence between 4 layer tracks. 
Each track quality is calculated by the evaluation function with full degree of freedom (4 and 12 parameters for blackness and linearity, respectively). By applying this method as third step, the S/N ratio on CS tracks is drastically improved by $\sim$100 times with only 1\% efficiency loss. Consequently fake tracks reduced to $\sim$10 tracks on CS. Furthermore, if there are tracks with highly ranking score, real tracks are automatically found and no eye check is performed.

\section{Tau neutrino analysis}

The four main decay channels of the ${\tau}$ lepton are ${\tau}$ ${\rightarrow}$ ${e}$ (17.4$\%$), ${\mu}$ (17.9$\%$), 1 hadron (49.5$\%$) and 3 hadrons (15.2$\%$). They are topologically classified as "kink" or "trident" if they have one or 3 charged daughters. Due to the similar lifetime and decay topologies, the charmed particles detection rate represents a benchmark of the ${\tau}$ detection efficiency \cite{bib20}. The decay products are identified by their large impact parameter (IP) with respect to the primary vertex. If the parent particle traverses at least one emulsion layer, the trajectory of both the parent and the decay daughter are clearly observed in the emulsion films.

Once a decay candidate event with large IP and without a tagged ${\mu}$/${e}$ at primary vertex is observed, a kinematical analysis is performed. The momenta of charged particles determined by multiple Coulomb scattering in the ECC, and by the information provided by the magnetic spectrometers and the total energy deposited in the instrumented target, acting as a calorimeter, are used for this analysis. The kinematical selection criteria used to identify ${\nu_{\tau}}$CC interactions are shown in table.\ref{tab:sel}. A search for nuclear fragments is performed both upstream and downstream of secondary vertex up to $|$tan${\theta}|$ = 3.0 with newly developed analysis technique \cite{bib21} to significantly reduce the background due to hadronic interactions, as described the details at next section.

\begin{table}[hbt]
  \caption{Kinematical selection}
  \label{tab:sel}
  \begin{tabular}{lcccc}
    \hline variable  & ${\tau}$ ${\rightarrow}$ 1${h}$ & ${\tau}$ ${\rightarrow}$ 3${h}$ & ${\tau}$ ${\rightarrow}$ ${\mu}$ & ${\tau}$ ${\rightarrow}$ ${e}$    \\
    \hline
      lepton-tag     & \multicolumn{4}{c}{No ${\mu}$ or ${e}$ at the primary vertex} \\
      ${z_{dec}}$ (${\mu}$m) & [44, 2600]  & $<$ 2600 & [44, 2600] & $<$ 2600 \\
      ${p^{miss}_T}$ (GeV/c) & $<$ 1 & $<$ 1 & / & / \\
      ${\Delta}{\phi}_{{\tau}H}$ (rad) & ${> \pi}$/2 & ${> \pi}$/2 & / & / \\
      ${p^{2ry}_T}$ (GeV/c) & $>$ 0.6(0.3) & / & $>$ 0.25 & $>$ 0.1 \\
      ${p^{2ry}}$ (GeV/c) & $>$ 2 & $>$ 3 & $>$ 1 and $<$ 15 & $>$ 1 and $<$ 15 \\
      ${\theta_{kink}}$ (mrad) & $>$ 20 & $<$ 500 & $>$ 20 & $>$ 20 \\
      $m, m_{min}$ (GeV/c$^2$) & / & $>$ 0.5 and $<$ 2 & / & / \\
    \hline
  \end{tabular}
\end{table}

\begin{figure}[ht]
\begin{center}
\includegraphics[clip, width=13.0cm]{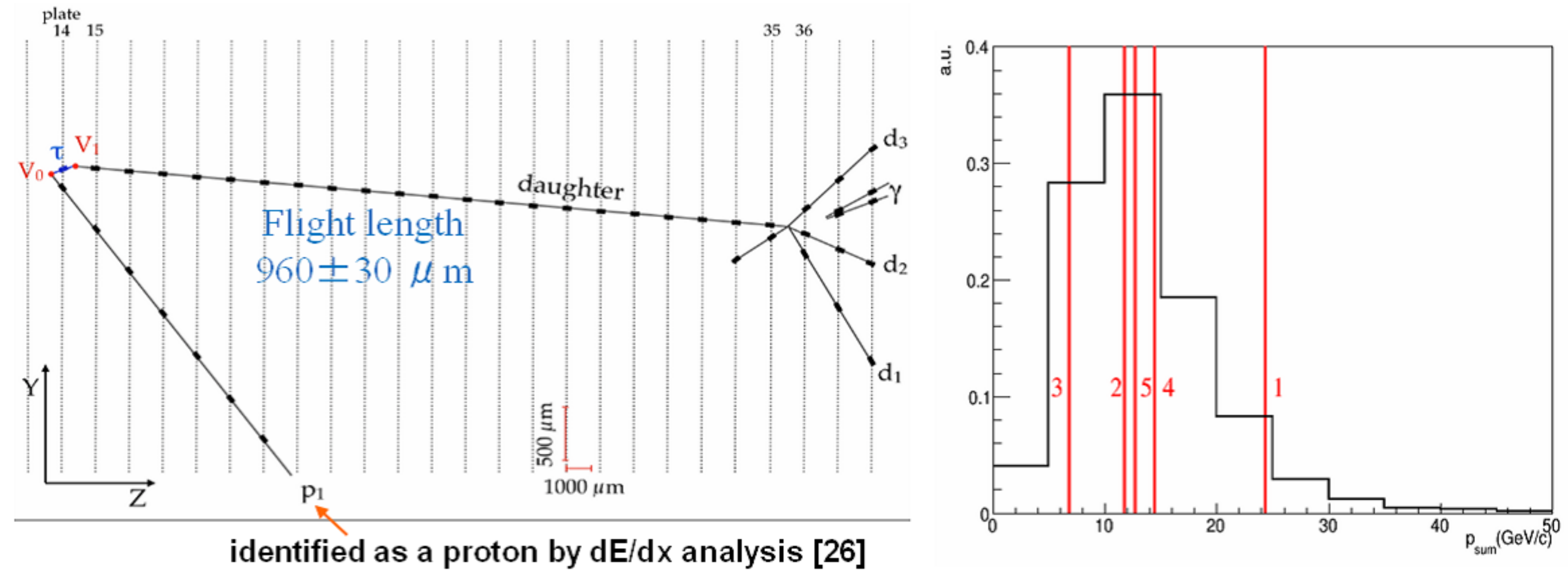}
\caption{The interaction vertex of the 5th ${\nu_{\tau}}$ event in the ECC (left). (right) Data-MC comparison of the scalar sum of the momenta of all particles measured in the ECC for five ${\nu_{\tau}}$ candidate events.}
\label{fig-2}
\end{center}
\end{figure}

The data sample of 1st and 2nd most probable bricks for all runs was analyzed to search for ${\nu_{\tau}}$ events. Five ${\nu_{\tau}}$ candidate events are observed in a sample of 5408 neutrino events \cite{bib7,bib22,bib23,bib24,bib25}. The fifth ${\nu_{\tau}}$ event is shown on the left of fig.\ref{fig-2}. One event is a ${\tau}$ ${\rightarrow}$ ${\mu}$ decay channel and the other four events are ${\tau}$ ${\rightarrow}$ ${h}$ or 3 ${h}$ decay channels. All tau parent tracks are recorded in an emulsion layer and the angle ${\Delta}{\phi}_{{\tau}H}$ in the transverse plane between the parent direction and the hadronic shower direction along the beam direction in all events is concentrated to the reverse direction (back to back). In the first ${\nu_{\tau}}$ event, the invariant mass of the charged decay daughter, assumed to be a ${\pi^-}$, and the two ${\gamma}$-rays is 640${^{+125}_{-80}}$(stat.)${^{+100}_{-90}}$(syst.) MeV/c${^2}$, which is compatible with the ${\rho}$(770) mass. In the second and the third ${\nu_{\tau}}$ event, the tau decays occurred in the plastic base. In the third ${\nu_{\tau}}$ event, the charge measurement of the muon daughter is performed using the track bending in the magnetized arm of the muon spectrometer. The muon charge is determined to be negative at 5.6 ${\sigma}$ significance. The total transverse momentum (${P_T}$) of the decay daughter is large for all kink events and no nuclear fragments is associated at the decay vertex in all the ${\nu_{\tau}}$ events. The scalar sum of the momenta of all particles measured in the ECC, ${p_{sum}}$, measured for the five events are compatible with the Monte Carlo distribution as shown on the right of fig.\ref{fig-2}. 
%These facts ensured that the observed ${\nu_{\tau}}$ candidates are real ${\nu_{\tau}}$ events with  highly reliability.

\section{Backgrounds}

\subsection{Main sources of background} \mbox{} \\
The three main sources of background for the ${\nu_{\tau}}$ appearance search are charmed particle decays, hadronic interactions and large-angle muon scattering (LAS). The corresponding contributions are estimated by simulation studies validated with real data \cite{bib20,bib27,bib28}.
\setlength{\leftmargini}{10pt}
\begin{itemize}
\item Charmed particles have similar masses and lifetimes as those of the ${\tau}$ lepton. If a muon at the primary vertex is not identified, a charm production and decay event represent a background. The charm background is estimated by a Monte Carlo tuned on CHORUS data and the uncertainty has been estimated to 20\%. This includes a contribution from the experimental uncertainty on the charm cross section (8\%), the hadronization fraction(10\%), and the statistical error of the OPERA charm control sample (15\%) which is used to validate decay detection efficiency.
\item The hadronic interactions become a background for the ${\tau}$ lepton decay in case the hadrons originated from the primary vertex in ${\nu_{\mu}}$NC event immediately interact. The hadronic background has been estimated by a FLUKA-based MC simulation benchmarked on systematic measurements of pion interactions in the OPERA ECC bricks. A good agreement between data from the complemental beam experiments and simulation is obtained and the uncertainty on hadronic background has also been estimated to 30\%. 
\item The estimation of the LAS background for ${\tau}$ ${\rightarrow}$ ${\mu}$ decay channel has been given by a GEANT4-base simulation take into account the effect of the nuclear form factor at the involved transferred momenta. The simulation has been validated with data in literature.
\end{itemize}

\subsection{Additional background reduction} \mbox{} Large angle track scanning\\
The hadronic background can be reduced by detecting nuclear fragments from a secondary vertex, since the existence of nuclear fragments is clear proof of a hadron interaction, not ${\tau}$ decay. As the nuclear fragments are emitted almost isotropically, a new automatic emulsion scanning system having a much wider angular tracking acceptance was developed to detect nuclear fragments efficiently \cite{bib21}. The new system is able to recognize tracks up to $|$tan${\theta}|$ = 3.5 (where ${\theta}$ is the track angle with respect to the perpendicular to the emulsion films), while conventional systems have an angular acceptance limited to $|$tan${\theta}|$ = 0.6. Nuclear fragments search from hadron interactions performed with this system is reported in \cite{bib27}. The same procedure has been applied to all the ${\nu_{\tau}}$ candidate events in OPERA checking for the presence of nuclear fragments at the decay vertex. A reduction of 30\% of the hadronic background was achieved. The tracking efficiency at large angle is high also for minimum ionizing particles, therefore the procedure could be applied to confirm decay topology of ${\nu_{\tau}}$ candidate events for wide-angle acceptance \cite{bib29}.

\section{Physics results}

\subsection{${\nu_{\mu}}$ ${\rightarrow}$ ${\nu_{\tau}}$ oscillation} \mbox{} \\
The CNGS beam run for five years, from 2008 till the end of 2012. An event sample corresponding to 17.97 ${\times}$ 10$^{19}$ pot has been recorded by OPERA. The data sample for the ${\nu_{\tau}}$ appearance search amounts to 5408; the corresponding expected ${\nu_{\tau}}$ signal (background) is of 2.64${\pm}$0.53 (0.25${\pm}$0.05) events, assuming ${\Delta}$\it{m}${^2_{23} = 2.44}$ ${\times}$ $10^{-3}$ \rm{eV}${^2}$, sin${^22}$ ${\theta_{23}}$ = 1.0. The systematic uncertainties are 20\% on the signal, 20\% on the charm background, 30\% on the hadronic background, and 50\% on the LAS background. Five ${\nu_{\tau}}$ candidate events have been observed: the first in the 2009 run data (${\tau}$ ${\rightarrow}$ ${h}$ decay channel), the second in the 2011 run data (${\tau}$ ${\rightarrow}$ 3${h}$ decay channel), the third in the 2012 run data (${\tau}$ ${\rightarrow}$ ${\mu}$ decay channel), the fourth in the 2012 run data (${\tau}$ ${\rightarrow}$ ${h}$ decay channel), and the fifth in the 2012 run data (${\tau}$ ${\rightarrow}$ ${h}$ decay channel).

The significance of the observation of the five ${\nu_{\tau}}$ candidate events is estimated as the probability for excluding the null hypothesis. A hypothesis test is based on the Fisher's method. The ${p_i}$ value of each individual channel are combined into an estimator 
${p^* = \prod_i p_i}$ for the background-only hypothesis. Each individual channel is calculated as the integral of the Poisson distribution for values larger or equal to the observed number of candidates. A background fluctuation probability (one-side) of 1.1 ${\times}$ ${10^{-7}}$ is obtained by comparing the observed ${p^*_{data}}$ with the sampling distribution of ${p^*}$. This corresponds to a significance of 5.1 ${\sigma}$ standard deviations. Another hypothesis test employing a profile-likelihood approach was also performed and a significance of 5.1 standard deviations was obtained \cite{bib25}.% This means "Discovery of ${\nu_{\tau}}$ appearance in a ${\nu_{\mu}}$ beam with the OPERA experiment".

The 90\% C.L. interval for ${{\Delta}\it{m}\rm{^2_{23}}}$ has been estimated with three different approaches using the profile likelihood ratio, the Feldman-Cousins method, and Bayesian statistics. Assuming full mixing, the best fit yields ${\Delta}{\it{m}\rm{^2_{23}}}$ = 3.3 ${\times}$ $10^{-3}$ eV$^2$, with a 90\% C.L. interval of [2.0, 5.0] ${\times}$ $10^{-3}$ eV$^2$, the difference among the three methods being negligible.

\subsection{${\nu_{\mu}}$ ${\rightarrow}$ ${\nu_e}$ oscillation and sterile neutrinos} \mbox{} \\
Thanks to the capability of electron identification and to the small contamination of ${\nu_e}$ in the CNGS neutrino beam, OPERA can also perform a ${\nu_{\mu}}$ ${\rightarrow}$ ${\nu_e}$ oscillation search. 
%Since the beam originally contains a small fraction of the ${\nu_e}$ and anti-${\nu_e}$, the deviation from expected number of ${\nu_e}$ in dependence of their energy will indicate the oscillation.
The procedure for ${\nu_e}$ analysis is described in \cite{bib30}. A systematic ${\nu_e}$ appearance search was applied to the data sample collected in 2008-2009 (5255 registered triggers), with 2853 events having a located and reconstructed vertex in ECC bricks. Out of 505 NC-like located events, 19 ${\nu_e}$ candidate events were observed with an expectation of 19.8${\pm}$2.8 (syst) events from beam contamination and 1.4 event from oscillation. In order to increase the signal to background ratio, candidates with an energy lower than 20 GeV were selected. By this requirement, 4.6 events are expected in the standard three-flavor mixing scheme while 4 are observed. An upper limit of sin${^22}$ ${\theta_{13}}$ $<$ 0.44 (90\% C.L.) is derived \cite{bib30}. Currently the ${\nu_e}$ event sample is about 2.5 times larger.

\begin{figure}[hb]
\begin{center}
\includegraphics[clip, width=13.0cm]{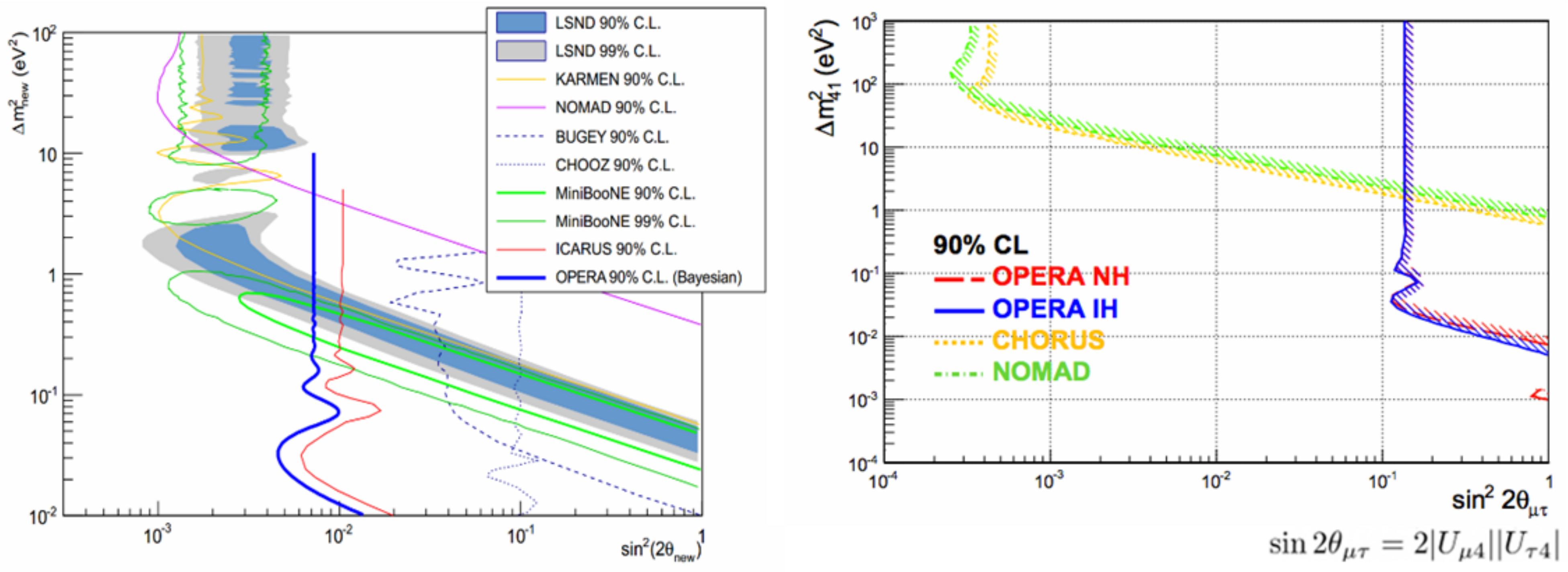}
\caption{Exclusion plots for the parameters of the non-standard ${\nu_{\mu}}$ ${\rightarrow}$ ${\nu_e}$ (left) and ${\nu_{\mu}}$ ${\rightarrow}$ ${\nu_{\tau}}$ (right) oscillations searched for by OPERA.}
\label{fig-3}
\end{center}
\end{figure}

The OPERA ${\nu_e}$ and ${\nu_{\tau}}$ results are used to derive limits on the mixing parameters of a massive sterile neutrino at the mass scale suggested by the LSND and MiniBooNE experiments \cite{bib31,bib32}. In ${\nu_{\mu}}$ ${\rightarrow}$ ${\nu_e}$ channel, an energy cut E $<$ 30 GeV is applied; 6 events are observed with an expectation of 9.4${\pm}$1.3 (syst) events, including ${\nu_e}$ events via three flavor oscillations. Based on a Bayesian statistical method, an upper limit and a sensitivity on sin${^22}$ ${\theta_{new}}$ for large ${\Delta}{\it{m}\rm{^2}\it{_{new}}}$ \rm{of} 7.2 ${\times}$ ${10^{-3}}$ and 10.4 ${\times}$ ${10^{-3}}$, respectively, are derived at 90\% C.L. (left panel of fig.\ref{fig-3}) \cite{bib30}. The OPERA ${\nu_{\tau}}$ events are also used to derive limits on the mixing parameter of a massive sterile neutrino in the 3$+$1 neutrino model. 
%Data sample for the analysis is about 75\% of the total statistics with four ${\nu_{\tau}}$ detection. 
The details are described in \cite{bib33}. At high ${\Delta}{\it{m}\rm{^2_{41}}}$ values, the measured 90\% C.L. upper limit on the mixing term sin${^22}$ ${\theta_{\mu\tau}}$ = 4${|U_{\mu4}|^2|U_{\tau4}|^2}$ is 0.116, independently of the mass hierarchy of the three standard neutrinos. The limits on ${\Delta}{\it{m}\rm{^2_{41}}}$ in the ${\nu_{\mu}}$ ${\rightarrow}$ ${\nu_{\tau}}$ appearance mode is extended down to values of ${10^{-2}}$ eV$^2$ at large mixing for sin${^22}$ ${\theta_{\mu\tau}}$ ${\gtrsim}$ 0.5 (right panel of fig.\ref{fig-3}) \cite{bib33}. The analysis is being updated using all 5 ${\nu_{\tau}}$ candidates. A preliminary result yields sin${^22}$ ${\theta_{\mu\tau}}$ $<$ 0.11 at 90\% C.L. \cite{bib34}.

\section{Conclusions}
The OPERA experiment on the CNGS beam has been taking data from 2008 till the end of 2012, collecting neutrino interaction events corresponding to 17.97 ${\times}$ 10$^{19}$ pot. Neutrino event search and reconstruction has been successfully carried out with newly developed emulsion analysis techniques. High speed scanning, precise alignment, high S/N discrimination and large angle scanning were performed. Five ${\nu_{\tau}}$ candidate events have been detected with the estimated background of 0.25 events, achieving the "Discovery of ${\nu_{\tau}}$ appearance in a ${\nu_{\mu}}$ beam" with 5.1 ${\sigma}$ significance.

The oscillation parameters for the ${\nu_{\mu}}$ ${\rightarrow}$ ${\nu_{\tau}}$ appearance has been determined as ${\Delta}{\it{m}\rm{^2_{23}}}$ = 3.3 ${\times}$ $10^{-3}$ eV$^2$; the 90\% C.L. interval is [2.0, 5.0] ${\times}$ $10^{-3}$ eV$^2$, assuming full mixing. The observed number of ${\nu_e}$ interactions is compatible with the non-oscillation hypothesis, allowing OPERA to set an upper limit in the parameter space available for a non-standard ${\nu_e}$ appearance. Limits on the mixing parameters of a massive sterile neutrino have also been derived and the exclusion limits on ${\Delta}{\it{m}\rm{^2_{41}}}$ in the ${\nu_{\mu}}$ ${\rightarrow}$ ${\nu_{\tau}}$ appearance channel has been extended down to values of ${10^{-2}}$ eV$^2$ at large mixing for sin${^22}$ ${\theta_{\mu\tau}}$ ${\gtrsim}$ 0.5.

\acknowledgments
This work was partially supported by JSPS KAKENHI Grant Numbers JP08J07061, JP23740184, JP25707019.

\end{document}